\documentclass[prx,twocolumn,floatfix,superscriptaddress,amsmath,caps,longbibliography]{revtex4-2}
\pdfoutput=1
\usepackage{dcolumn,graphicx,color,booktabs,microtype,afterpage} 
\makeatletter\renewcommand{\fnum@figure}[1]{\figurename~\thefigure.}\makeatother
\makeatletter\renewcommand{\fnum@table}[1]{\tablename~\thetable.}\makeatother
\usepackage[colorlinks,plainpages=false,linkcolor=blue,urlcolor=blue,citecolor=blue,pdfpagemode=UseNone,pdfstartview=FitBH]{hyperref}
\usepackage{bm} % bold math
\usepackage[T2A]{fontenc}

\newcommand{\HFO}{HoFeO$_3$}
\newcommand{\HFAO}{HoFe$_{1-x}$Al$_x$O$_3$}

\newcommand{\HFAOXX}{HoFe$_{0.8}$Al$_{0.2}$O$_3$}

\newcommand{\K}{K}
\newcommand{\M}{$\mu_\mathrm{B}/\mathrm{f.u.}$}
\newcommand{\Ho}{Ho$^{3+}$}
\newcommand{\Fe}{Fe$^{3+}$}

\begin{document}
	
\title{Effect of Magnetic Vacancies on the Spontaneous Spin-Reorientation Transition in \HFAO\ Single Crystals}

\author{S. A. Skorobogatov}
\thanks{Corresponding author: saskorobogatov@iph.krasn.ru}
\affiliation{Kirensky Institute of Physics, Federal Research Center KSC SB RAS, Krasnoyarsk, 660036 Russia}
\author{V. A. Babkina}
\affiliation{Kirensky Institute of Physics, Federal Research Center KSC SB RAS, Krasnoyarsk, 660036 Russia}
\author{M. S. Pavlovskii}
\affiliation{Kirensky Institute of Physics, Federal Research Center KSC SB RAS, Krasnoyarsk, 660036 Russia}
\affiliation{Siberian Federal University, Krasnoyarsk 660041, Russia}
\author{M. I. Kolkov}
\affiliation{Petersburg Nuclear Physics Institute named by B.P.Konstantinov of NRC «Kurchatov Institute», 188300, Russia, Leningradskaya Oblast, Gatchina, 188300, Russia}
\author{S. V. Semenov}
\affiliation{Kirensky Institute of Physics, Federal Research Center KSC SB RAS, Krasnoyarsk, 660036 Russia}
\affiliation{Siberian Federal University, Krasnoyarsk 660041, Russia}
\author{I. N. Khoroshiy}
\affiliation{Kirensky Institute of Physics, Federal Research Center KSC SB RAS, Krasnoyarsk, 660036 Russia}
\author{S. E. Nikitin}
\affiliation{Kirensky Institute of Physics, Federal Research Center KSC SB RAS, Krasnoyarsk, 660036 Russia}
\author{K. A. Shaykhutdinov}
\affiliation{Kirensky Institute of Physics, Federal Research Center KSC SB RAS, Krasnoyarsk, 660036 Russia}

\begin{abstract}

In this work, we report the first growth of single crystals of the substitution series \HFAO\ with aluminium concentrations up to $x=0.2$ and investigate the evolution of their spontaneous spin-reorientation transition (SRT). Among rare-earth orthoferrites, \HFO\ exhibits a distinctive sequence of magnetic phases ($\Gamma_4$–$\Gamma_{24}$–$\Gamma_{12}$–$\Gamma_2$). This complex sequence arises from the competition between the $K_{ac}$ and $K_{ab}$ anisotropies associated with the \Ho\ ions and the effective magnetic field produced by the weak ferromagnetic moment of the canted \Fe\ sublattice. 
Introducing magnetic vacancies perturbs the antiferromagnetic compensation of the \Fe\ subsystem in the $ab$ plane and generates an additional effective magnetic field acting on the \Ho\ ions. This field alters the balance between the $K_{ac}$ and $K_{ab}$ anisotropies within the SRT temperature range and thereby broadens the stability range of the $\Gamma_{12}$ phase in the magnetic phase diagram.

\end{abstract}

\maketitle{}

\section{Introduction}

Rare earth orthoferrites with the general formula \textit{R}FeO$_3$ are an interesting family of oxide crystals (space group No. 62, setting \textit{Рbnm}) from the viewpoint of magnetic properties. They possess a weak ferromagnetic moment caused by the canting of the antiferromagnetic sublattices, which is induced by the antisymmetric Dzyaloshinskii–Moriya interaction~\cite{dzyaloshinsky1958thermodynamic, moriya1960anisotropic}. They exhibit a variety of magnetic phenomena, including spontaneous spin-reorientation transitions (SRT)~\cite{belov1974new, belov1979orientational, white1969review}, ultrafast magnetization control by femtosecond laser pulses~\cite{kimel2005ultrafast, de2011laser}, nontrivial spin dynamics in the rare earth subsystem~\cite{Nikitin2018} to mention a few. Many of these effects have considerable potential for practical applications. Undoubtedly, the most studied among the above mentioned phenomena occurring in orthoferrites is the spontaneous spin reorientation transition, which was discovered and thoroughly investigated more than half a century ago\cite{wolfe1967temperature}. The temperature of this transition varies over a wide temperature range – from a few kelvin for YbFeO$_3$~\cite{Nikitin2018} up to 450~K for SmFeO$_3$~\cite{kang2017spin} – and is determined by the interaction between the iron and rare earth subsystems. To change the SRT temperature, isovalent substitution in the iron sublattice is often used~\cite{belov1979orientational, fan2022thermal, su2019spin, sun2023pr}, which allows one to shift the SRT even to room temperatures.

In our recent works devoted to the growth and study of a series of HoFe$_{1-x}$Mn$_x$O$_3$ single crystals (0 < $x$ < 1)~\cite{shaihutdiniov2024control, shaihutdiniov2024magnet, knyazev2024effect}, it was shown that substitution of iron by manganese significantly increases the SRT temperature. For example, at a manganese concentration $x = 0.4$, the transition temperature was 294~K, while for the unsubstituted compound \HFO\ it is 58~K. Moreover, such substitution changes the nature of the magnetic phase transition from type II ($\Gamma_4 \rightarrow \Gamma_2$) at $x=0$, in which the weak ferromagnetic moment rotates from the \textbf{c} axis to the \textbf{a} axis upon cooling, to type I ($\Gamma_4 \rightarrow \Gamma_1$), for which the weak ferromagnetic moment vanishes in the $\Gamma_1$ phase.

As a natural continuation of these works, it would be interesting to study the magnetic properties of samples with nonmagnetic isovalent substitutions, such as Al$^{3+}$, Ga$^{3+}$, Sc$^{3+}$. 
According to the available literature on \HFAO\ single crystals, we note the study of Ref.~\cite{vorob1991orientational}, which investigated HoFe$_{0.97}$Al$_{0.03}$O$_3$ and HoFe$_{0.93}$Al$_{0.07}$O$_3$ single crystals.
These crystals were grown by spontaneous crystallization from a flux melt. As noted by the authors of Ref.~\cite{vorob1991orientational}, ``the quantitative content of Al$^{3+}$ ions was determined by X-ray spectral analysis with an accuracy of 4\%''. Such uncertainty is typical of the flux-growth method, because during spontaneous crystallization the actual concentration of the substituent ion in the crystal is not fixed by its nominal concentration in the starting mixture.

The main conclusion of these works is the discovery of a very strong influence of nonmagnetic impurity ions on the magnetic properties of rare‑earth orthoferrites. Thus, the results of Ref.~\cite{vorob1991orientational} suggest that this promising line of research on magnetic vacancies in orthoferrites has received little attention since the early 1990s. In our view, this may be largely due to the difficulty of growing high-quality single crystals.
Those works demonstrated a very strong effect of vacancies on the magnetic properties. Yet, the spontaneous crystallisation technique does not readily yield a series of single crystals with controlled substitutions. The optical floating‑zone method overcomes this challenge: single‑crystal growth takes place through recrystallisation of the initial polycrystalline material, with the reagent charge error kept to no more than 0.05\%. Accordingly, the main objective of this work is to study how magnetic vacancies affect the magnetic properties of orthoferrites, employing a series of high‑quality \HFAO\ samples obtained by the optical floating‑zone method.

\section{EXPERIMENTAL DETAILS} 

To prepare \HFAO\ samples with $x=0$, 0.05, 0.1, and 0.2, Ho$_2$O$_3$, Fe$_2$O$_3$, and Al$_2$O$_3$ powders (99.9\%, Alfa Aesar) were mixed in the required stoichiometric proportions and annealed at 800~$^\circ$C for 18~h.
The annealed powders were poured into a rubber mould and pressed in a hydrostatic press at a pressure of $\approx 100$ MPa. The obtained cylindrical samples were then annealed in a vertical furnace at 1450$^\circ C$ for 16 hours. After that, the synthesised polycrystalline \HFAO \ samples were placed into an optical floating‑zone furnace (FZ‑T‑4000‑H‑VIII‑VPO‑PC, Crystal Systems Corp.), where the single‑crystal growth was carried out. The growth was performed in air at ambient pressure and with a relative rod rotation speed of 30 rpm. The growth rate varied from 3 to 1 mm/h depending on the iron‑to‑aluminium ratio in \HFAO.

Powder X‑ray diffraction patterns of \HFAO \  were obtained at room temperature on a Rigaku SmartLab 3~kW X‑ray diffractometer with a point detector. The X‑ray radiation was Cu-K$\alpha$ with a K$_\beta$ filter. The scanning step size 2$\theta$ was 0.02$^\circ$, with a counting time of 10 second per step; measurements were performed at room temperature. For orientation of the obtained crystals along the crystallographic axes, a Laue diffractometer (Photonic Science) was used. 
The magnetization measurements were performed in the temperature range 4.2 – 350~K using PPMS‑6000 (Quantum Design) instrument. High‑temperature magnetization measurements (400 – 1000~K) were carried out on a LakeShore Cryotronics VSM 8604 vibrating‑sample magnetometer. After thermal cycling up to 1000~K, we verified the reproducibility of the characteristic temperatures, including the N'eel and spin-reorientation transition temperatures, as well as the magnetization values. All measured quantities were fully reproducible.

\section{RESULTS AND ANALYSIS}

We start presentation of our results with the structural analysis data. For the compositions with $x = 0.05$ and $0.2$, powder X‑ray diffraction was performed to refine the lattice parameters. The obtained lattice parameters are summarized in Table~\ref{crystal}. All lattice parameters along with cell volume show monotonic decrease with $x$, in accord with the data on two reference samples \HFO\ and HoAlO$_3$~\cite{shaihutdiniov2024control, hammann1977etude}.

\begingroup
\renewcommand{\arraystretch}{1.5}
\begin{table}[h]
\caption{ \ Lattice parameters and unit‑cell volume of  \HFAO \  single crystals.}
\vspace{5 pt}
\centering
\begin{tabular}{ c|c|c|c|c }

\hline 
 $x$ & a, \AA & b, \AA & c, \AA & V, \AA$^3$ \\ 
\hline
 0.00~\cite{shaihutdiniov2024control} \ & \ 5.2921 \ & \ 5.6029 \ & \ 7.6151 \ & \ 225.80 \ \\
 0.05 \ & \ 5.2701 \ & \ 5.5760 \ & \ 7.5862 \ & \ 222.93 \ \\
 0.20 \ & \ 5.2563 \ & \ 5.5429 \ & \ 7.5527 \ & \ 220.05 \ \\
 1.00~\cite{hammann1977etude} \ & \ 5.1820 \ & \ 5.3240 \ & \ 7.3700 \ & \ 203.33 \ \\
\hline

\end{tabular} 
\label{crystal}
\end{table}
\endgroup

\begin{figure}[b]
\centering\includegraphics[width=0.6\columnwidth]{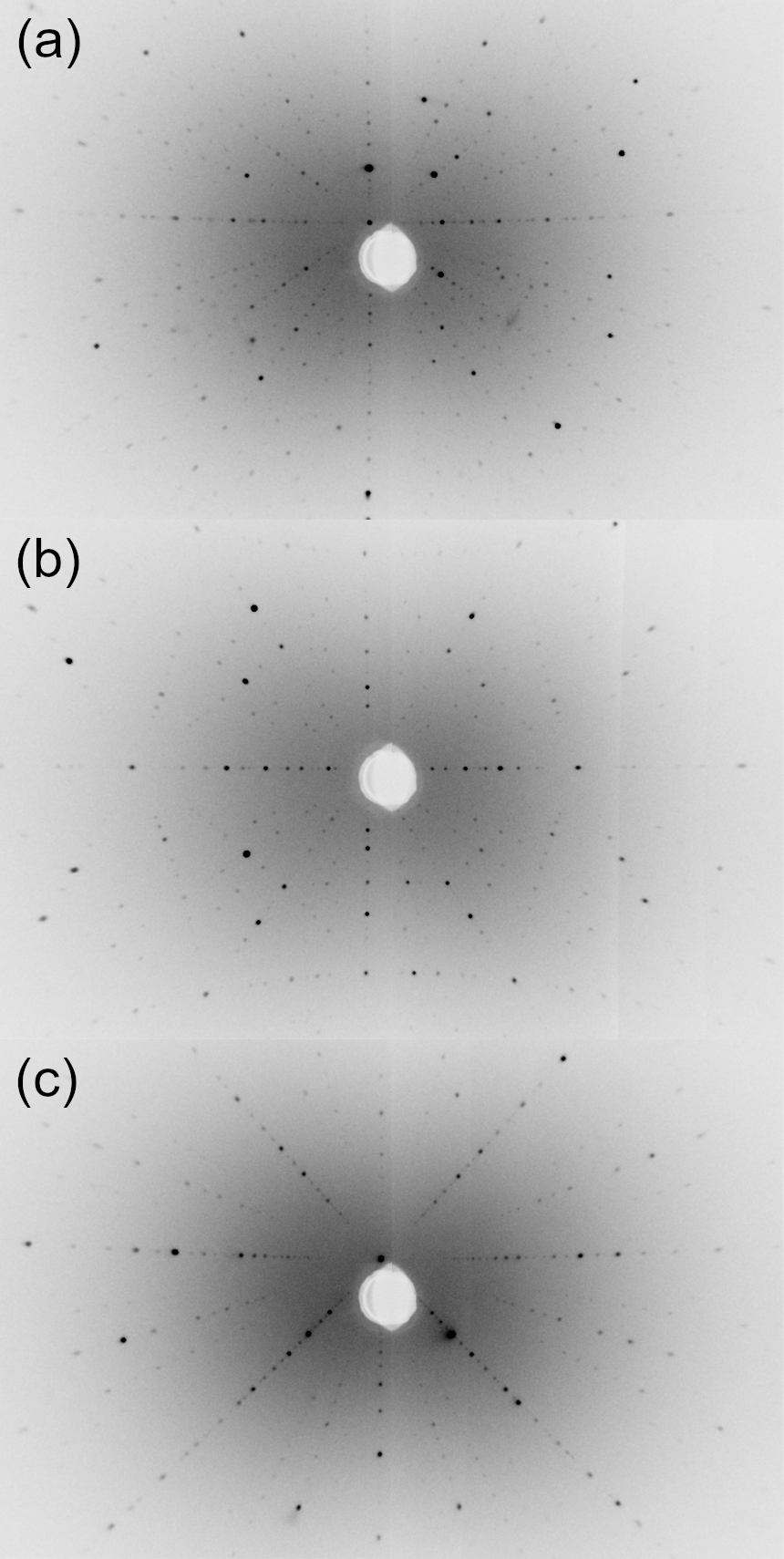}	
    \caption{ Laue patterns of the \HFAOXX  \ single crystal from the (a) (100), (b) (010), and (c) (001) reflection planes.}
    \label{Laue} 
\end{figure}

Figure\ref{Laue} presents the Laue diffraction pattern of \HFAOXX \ recorded from the (100), (010), and (001) crystallographic planes. The high quality of these diffraction patterns demonstrates a high quality of the single crystal samples used in this work.

\begin{figure*}[t]
\centering\includegraphics[width=\textwidth]{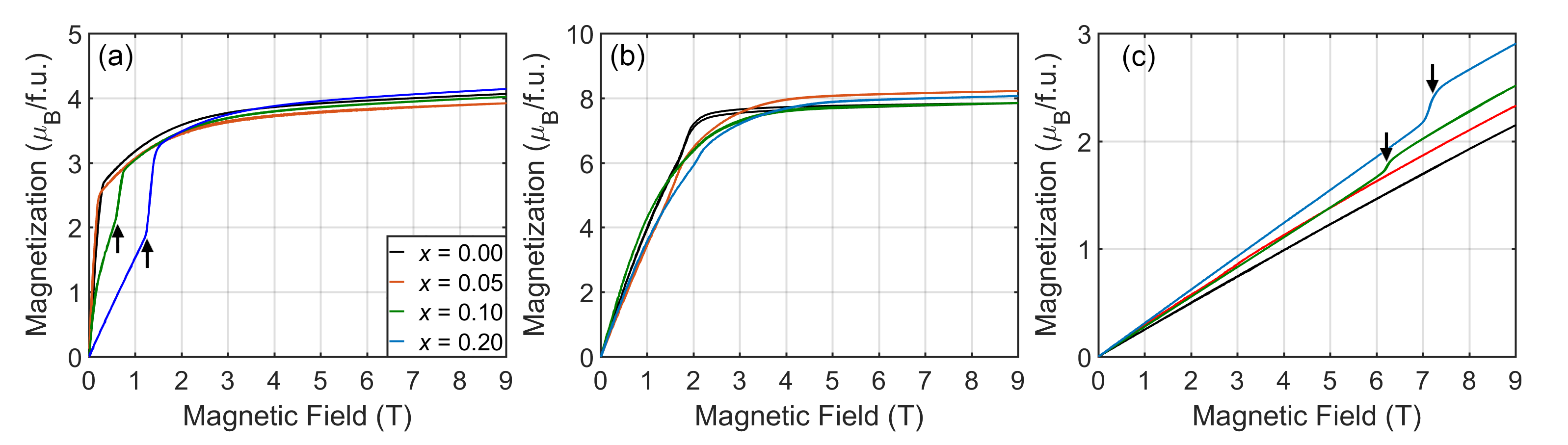}	
\caption{Field dependences of magnetization $M(H)$ for the studied substituted \HFAO\  compositions and the parent \HFO. Panels (a), (b), and (c) correspond to the crystallographic directions \textbf{a}, \textbf{b} and \textbf{c} at T = 4.2~\K. Arrows indicate spin‑flop transitions.}
\label{M(H)} 
\end{figure*}

In the studied compound, there are two magnetic subsystems associated with \Ho\ and \Fe\ ions. 
The \Fe\ subsystems orders antiferromagnetically and yields a weak anisotropy that stabilizes $\Gamma_4$ phase at high temperatures.
The \Ho\ subsystem does not order down to at least 4.2~K and exhibits strong anisotropy due to crystal electric field effect. 
The strength of this anisotropy is increasing with temperature decrease.
The SRT arises from the interaction of the two subsystems and is sensitive to changes occurring in both of them.

To reveal the influence of magnetic vacancies on the magnetic behavior of the studied compositions, we present the field dependences of magnetization $M(H)$ along the three crystallographic directions for all four crystal as shown in Fig.~\ref{M(H)}.
The anisotropic character of the \Ho\ ion can be seen clearly and is preserved for all substitution compositions. Based on our measurements performed at 9~T, where the $M(H)$ dependences are close to saturation for \textbf{a} and \textbf{b} directions, the angle between the \Ho \ magnetic moment and the \textbf{a} axis is $\alpha = \mathrm{arctg} (m_b/m_a) \simeq 62-65^\circ$. The magnetic moment value for the studied compositions lies in the range from 8.8 to 9.1~\M, which agrees with the results of~\cite{belov1979orientational, belov1974new} and is close to the value for the free \Ho\ ion (10 \M)~\cite{coey2010magnetism}. In orthoferrites, the \Ho\ ion can be regarded as an Ising ion~\cite{belov1979orientational}, reflecting the strong preferential orientation of the \Ho\ magnetic moments along a specific crystallographic direction and the applicability of the effective quasispin-$1/2$ approximation at low temperatures, $T < 10$~K. Within this approximation, only two states are available, corresponding to spin-up and spin-down orientations, with an abrupt transition between them. Consequently, the magnetization measured along any crystallographic direction is determined by the projections of the \Ho\ moments onto the measurement axis, resulting in a rapidly saturating $M(H)$ dependence. The presented results demonstrate the preservation of the Ising character of the \Ho \ ion behaviour in the substituted single crystals.

The $M(H)$ dependences also change systematically with increasing $x$: spin-flop transitions emerge for fields applied along the \textbf{a} and \textbf{c} axes, while the magnetization along the \textbf{c} axis increases.
We attribute the emergence of the spin-flop transitions and the changes in magnetization to the formation of the intermediate $\Gamma_{12}$ phase~\cite{vorob1991orientational}, as further supported below by the temperature-dependent magnetization $M(T)$ measurements.

Before discussing the $M(T)$ dependences, we briefly describe the measurement protocol. Because the weakly ferromagnetic state exhibits magnetic hysteresis, measurements in low fields below the coercive field require a specific field-history procedure to ensure reproducible magnetization values. Rather than approaching the target field from zero field, we first magnetized the sample in a field exceeding the coercive field and then reduced the field to the desired. This procedure ensured a single-domain state before recording the $M(T)$ dependence.

\begin{figure}[b]
\centering\includegraphics[width=\columnwidth]{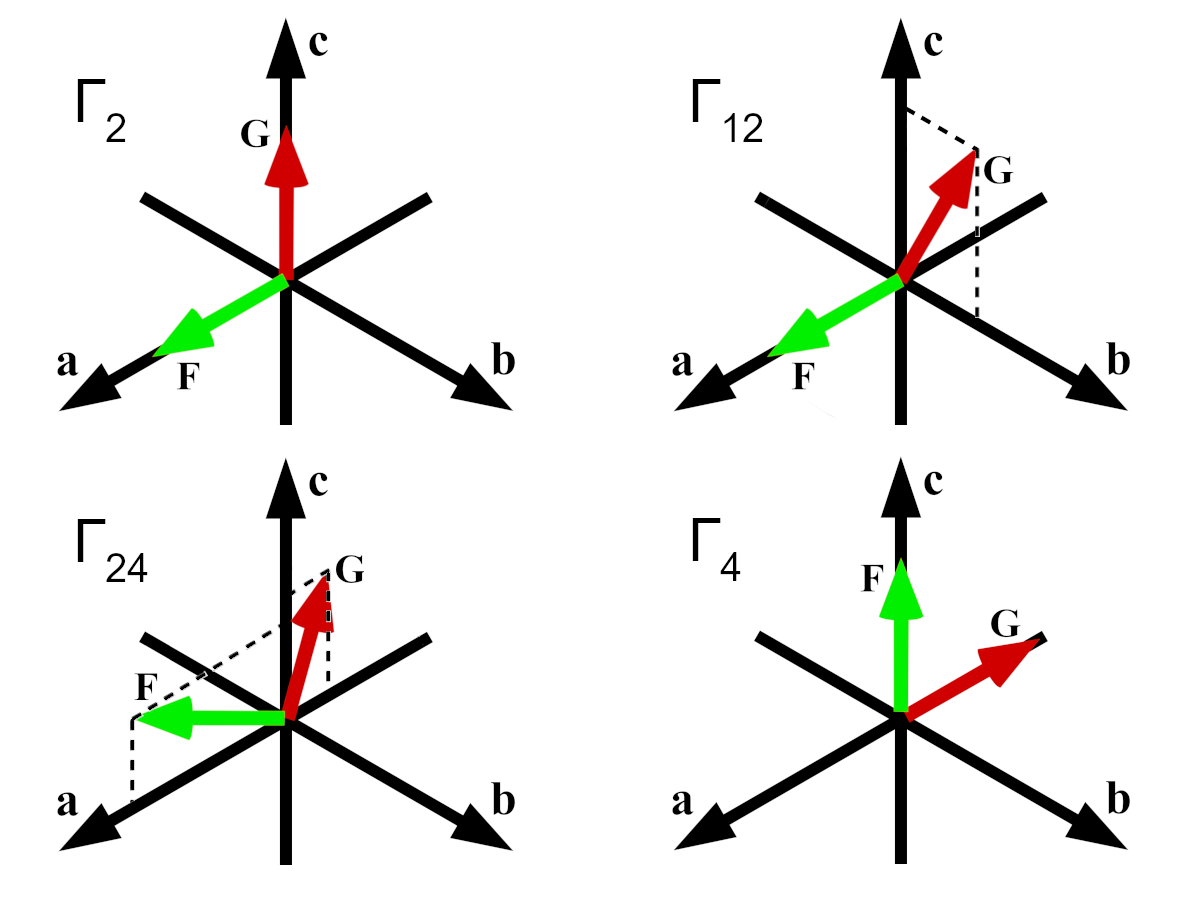}	
\caption{Schematic representation of the $\Gamma_2$, $\Gamma_{12}$, $\Gamma_{24}$ and $\Gamma_4$ phases. Black arrows show crystallographic directions in the $Pbnm$ setting, green arrow – ferromagnetic vector \textbf{F}, red arrow – antiferromagnetic vector \textbf{G}.}
\label{Gamma_phase} 
\end{figure}

When starting to discuss the experimentally observed SRT features, we use the standard notation for phases in the vector representation: $\Gamma_1$ = \textbf{A}$_x$\textbf{G}$_y$\textbf{C}$_z$, $\Gamma_2$ = \textbf{F}$_x$\textbf{C}$_y$\textbf{G}$_z$, $\Gamma_3$ = \textbf{C}$_x$\textbf{F}$_y$\textbf{A}$_z$ and $\Gamma_4$ = \textbf{G}$_x$\textbf{A}$_y$\textbf{F}$_z$, where \textbf{G}, \textbf{C} and \textbf{A} are antiferromagnetic vectors, and \textbf{F} is the ferromagnetic vector, see Fig.~\ref{Gamma_phase}. In the case of orthoferrites, the magnetic structure can be described using the pair of vectors \textbf{G} and \textbf{F}. 

\begin{figure}[t]
\centering\includegraphics[width=\columnwidth]{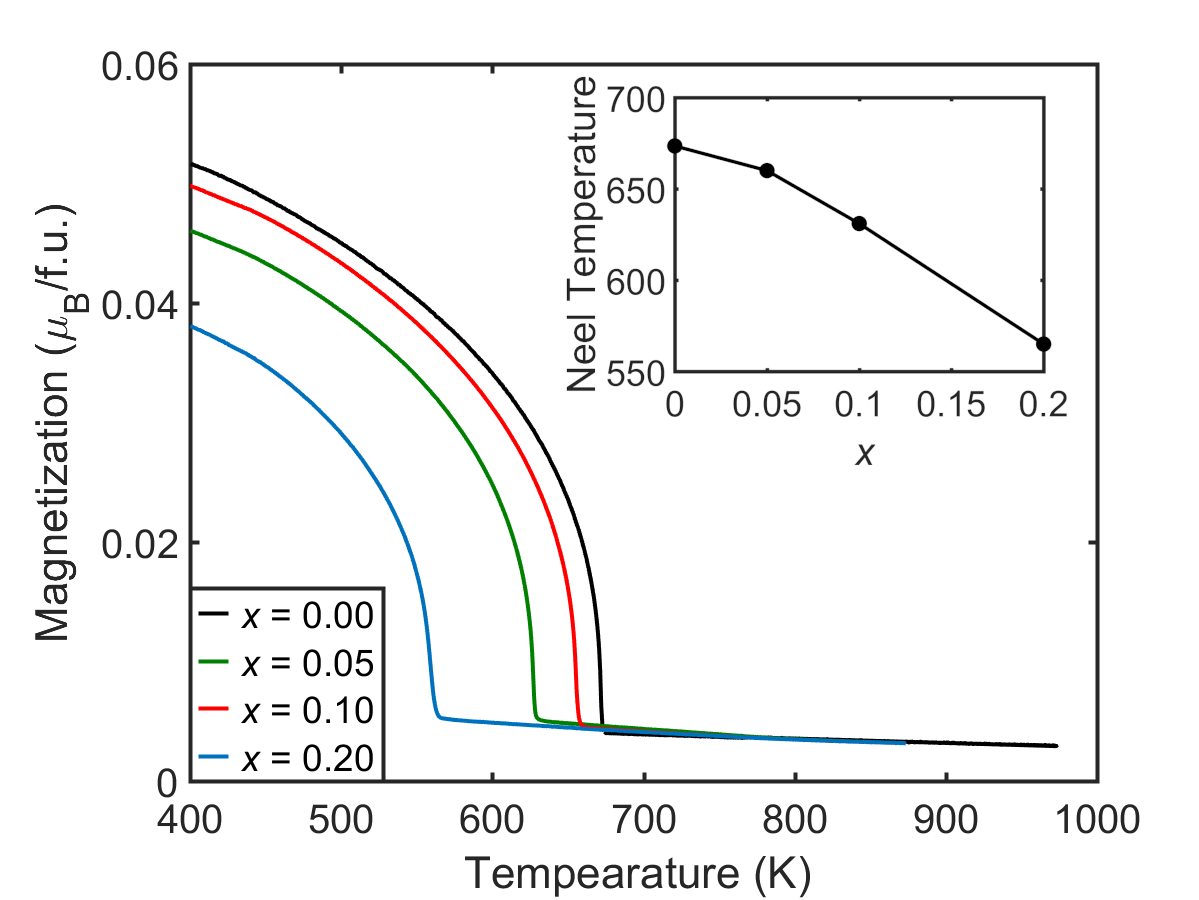}	
\caption{ $M(T)$ dependences of \HFAO \ obtained at an external magnetic field of 0.1~T in the temperature range 400 – 1000~K. Measurements were performed along the \textbf{c} direction. Inset: Néel temperature as a function of substitution level.}
\label{Neel} 	
\end{figure}

\begin{figure}[t]
\centering\includegraphics[width=\columnwidth]{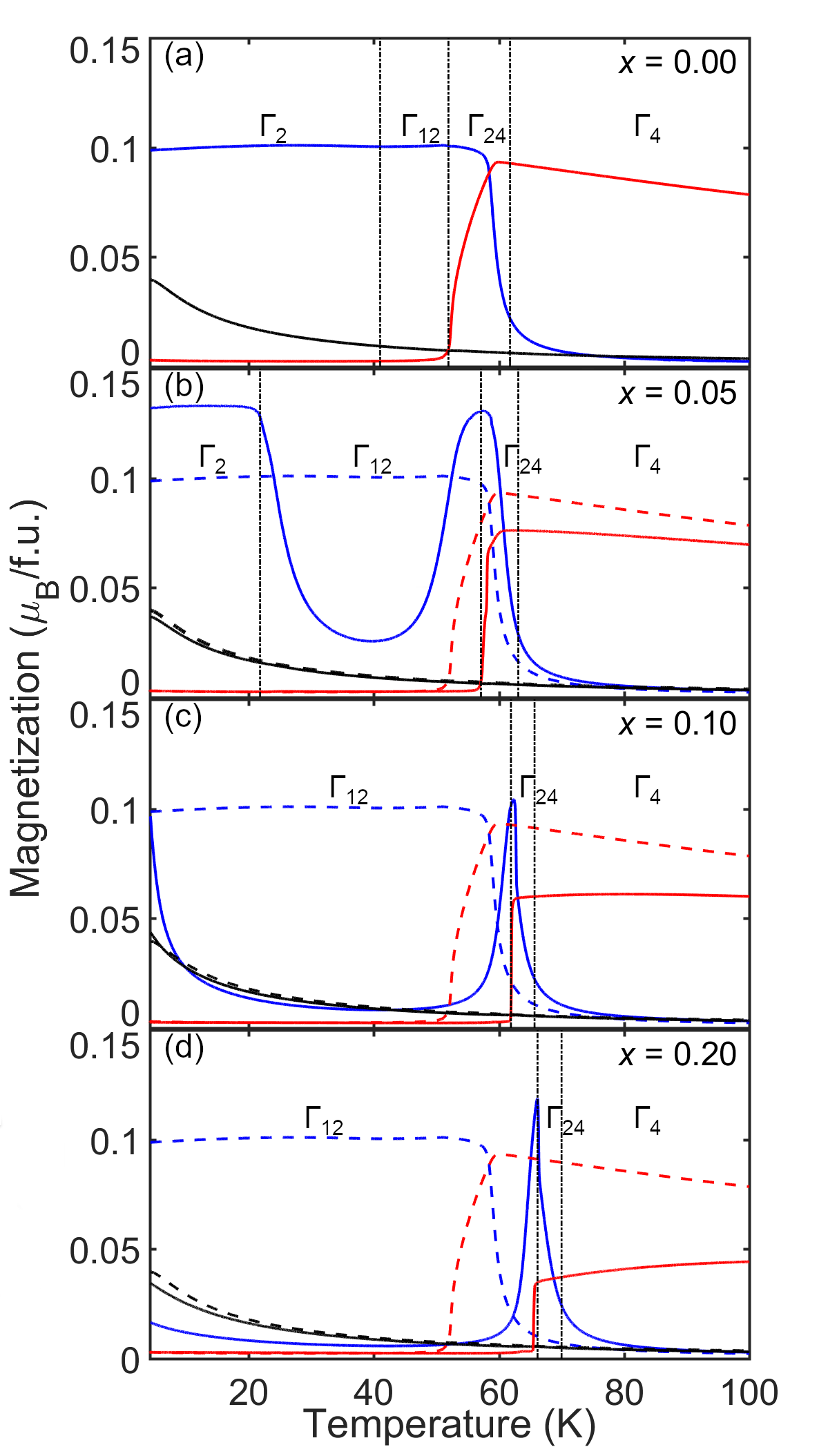}	
\caption{ $M(T)$ dependences of \HFAO \ obtained at an external magnetic field of 0.01~T. Colors indicate measurements along crystallographic directions: blue – \textbf{a}, black – \textbf{b}, red – \textbf{c}. Panels (a–d) show $M(T)$ for compositions with $x = 0.0, 0.05, 0.1, 0.2$. Solid lines correspond to substituted compositions, dashed lines show data for \HFO. Vertical black dash‑dotted lines indicate the SRT phase boundaries. For the crystallographic direction with the weak ferromagnetic moment, data obtained upon cooling are shown for the \textbf{c} direction, and upon heating for the \textbf{a} direction.}
\label{M(T)} 	
\end{figure}

Following the measurement protocol and notation introduced above, we now consider the $M(T)$ curves shown in Figs.~\ref{Neel} and~\ref{M(T)}, which compare all three substituted compositions with the parent \HFO. We first focus on the high-temperature $\Gamma_4$ phase to determine the N\'eel temperature. 

In Fig.~\ref{Neel}, two features should be noted. The first is the change in the N\'eel temperature with increasing concentration of magnetic vacancies. In this case, the magnetic order in the iron subsystem is disturbed, leading to a decrease in the total magnetic energy, and magnetic ordering occurs at lower temperatures. The second feature is the change in the spontaneous magnetization value. In Fig.~\ref{Neel}, a monotonic decrease in its value is observed. When discussing the reasons for this effect, one can assume that this is a consequence of the substitution of the original subsystem by nonmagnetic Al ions.

The discussion of the obtained $M(T)$ dependences should begin with the SRT in the parent \HFO. Based on work~\cite{vorob1989unusual}, where the results of thermal expansion, Young’s modulus measurements, etc. are examined in detail, the angle (15–20$^\circ$) of the deviation of the antiferromagnetic vector \textbf{G} from the \textbf{c} axis in the \textbf{bc} plane was determined. The authors of that work determined the temperature boundaries of the phases within the SRT, namely $\Gamma_4$ – $\Gamma_{24}$ $(\approx 60$ K), $\Gamma_{24}$ – $\Gamma_{12}$ ($\approx 50$ K)  and $\Gamma_{12}$ – $\Gamma_2$ ($\approx 40$ K). In our work, these phase boundaries are marked in Fig.~\ref{M(T)} a by vertical black dash‑dotted lines. The phase boundary temperatures were determined using inflection point (thin black dashed lines) in the region of appearance (for the \textbf{a} axis) and disappearance (for the \textbf{c} axis) of the weak ferromagnetic moment for the corresponding directions. The temperature of the $\Gamma_{12}$ – $\Gamma_2$ phase boundary was taken from~\cite{vorob1989unusual}.

Now we turn to the discussion of the $M(T)$ data obtained for different concentrations of Al as shown in Figs.\ref{M(T)}(b-d). Several trends can be seen. First, the SRT shifts to higher temperatures, approximately 2.5~K per 5\% aluminium. This is significantly smaller than the previously observed increase in the SRT temperature, about 40~K per 5\% manganese, for the o Fe$_{1-x}$Mn$_x$O$_3$ series~\cite{shaihutdiniov2024control}. Second, the $\Gamma_{24}$ phase is observed over a narrower temperature range. The decrease in the weak ferromagnetic moment with increasing $x$ is also clearly seen from the $M(T)$ along \textbf{c}. However, along the \textbf{a} axis, on the contrary, the magnetization is larger for substituted compositions compared to the parent one. Also, at the temperature of 40~K, indicated in~\cite{vorob1989unusual} as the boundary between $\Gamma_{12}$ and $\Gamma_2$, a minimum of magnetization along the \textbf{a} axis is observed for $x = 0.05$. The broadening of the temperature range of the minimum indicates an enlargement of the $\Gamma_{12}$ phase, which is does not manifest itself in magnetization measurements in the parent compound \HFO. 

\begin{figure}[t]
\centering\includegraphics[width=\columnwidth]{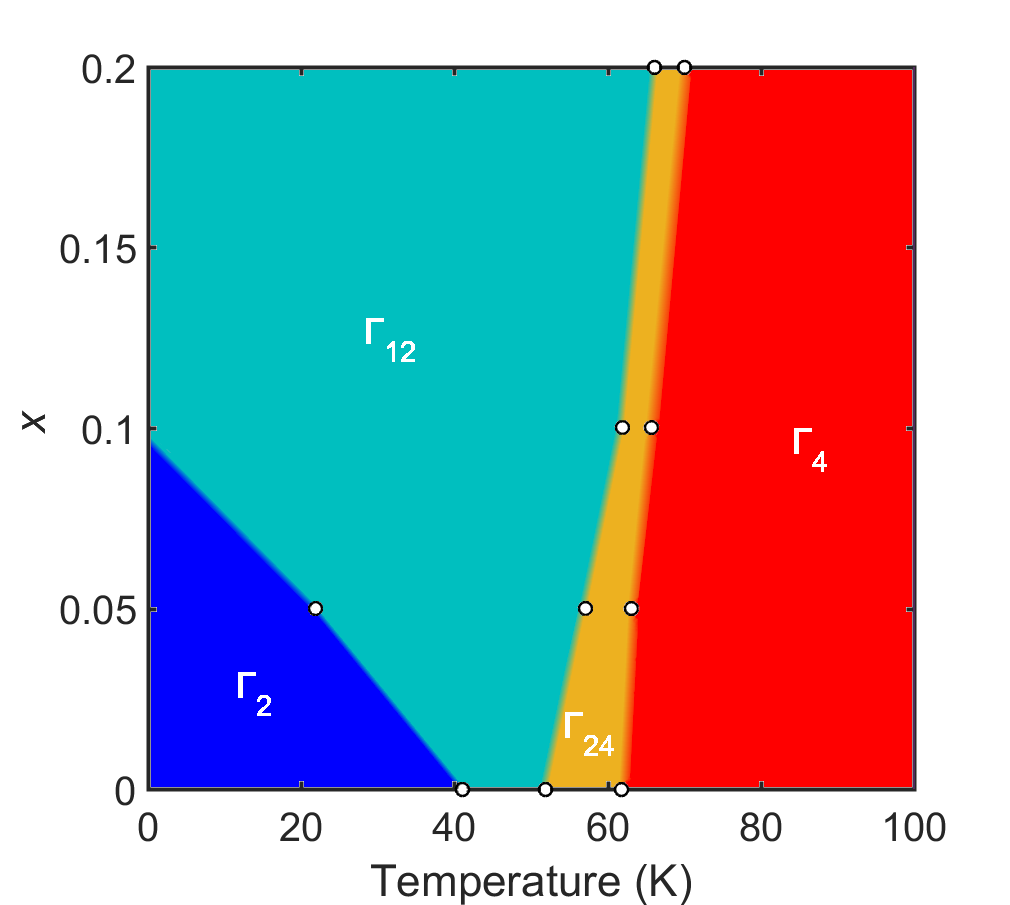}	
\caption{Phase diagram for the studied compositions $x$; colored areas represent different phases. Points mark the phase boundaries determined from the experimental $M(T)$ data.}
\label{phase} 
\end{figure}

From the $M(T)$ measurements, a phase diagram (Fig.~\ref{phase}) was constructed, which shows the concentration evolution of the magnetic phases $\Gamma_2$, $\Gamma_{12}$, $\Gamma_{24}$ and Г$_{4}$ close to the SRT.

\section{DISCUSSION}

In the $\Gamma_4$ phase of the parent compound \HFO\ all iron ion moments are predominantly oriented in the $ab$ plane with the antiferromagnetic vector \textbf{G} directed along the \textbf{a} axis. The spontaneous ferromagnetic moment observed experimentally is due to the canting of magnetic moments as a result of DMI. The direction of this moment is characterized by the ferromagnetic vector \textbf{F}. 
 According to~\cite{vorob1989unusual}, during the SRT the intermediate $\Gamma_{24}$ phase is stabilized, where the vector \textbf{F} smoothly rotates in the $ac$ plane towards the \textbf{a} direction, and the vector \textbf{G} towards the \textbf{c} direction. This process is associated with the influence of the magnetic anisotropy of the rare‑earth subsystem $K_{ac}$ as the temperature decreases. Near the SRT temperature, this contribution dominates over the iron anisotropy initially directed along \textbf{c} and over the second component of the \Ho\ anisotropy $K_{ab}$. At temperatures of 40–50~K a deviation of the \textbf{G} vector in the $bc$ plane is observed while the direction of the \textbf{F} vector remains along \textbf{a}~\cite{vorob1989unusual}. This $\Gamma_{12}$ transition cannot be seen in the magnetization measurements of \HFO, and its appearance is explained by the contribution of the $K_{ab}$ anisotropy, leading to a small (15–20$^\circ$) deviation of the \textbf{G} vector from the \textbf{c} axis in the $bc$ plane.

 The behavior of the SRT has changed upon the introduction of magnetic vacancies into the iron subsystem, as shown in Figs.~\ref{Neel} and~\ref{M(T)}. In the $\Gamma_4$ phase, a decrease in the magnetization magnitude is observed for all $x$, which is expected when iron is replaced by nonmagnetic aluminum. In the region where the intermediate phase $\Gamma_{24}$ exists, a sharp increase in the weak ferromagnetic moment is observed. The origin of this magnetization is the iron subsystem, since the rare‑earth subsystem is not ordered, as evidenced by the $M(H)$ dependences in Fig.~\ref{M(H)}. Thus, one can assume an increase in the canting angles of the iron moments when the vector \textbf{F} tends towards the \textbf{a} direction.
 
Since the crystal structure of \HFAO\  is a distorted perovskite, an iron ion can have seven configurations of the nearest environment in terms of the number of aluminum impurity ions. Thus, an iron ion with different probabilities can be surrounded by 0 to 6 aluminum ions. In solid solutions, the impurity is distributed statistically, and the probability of each configuration at a given concentration can be calculated as:

\begin{equation}
\label{eq:binom1}
    P(k) = C(n,k) \cdot x^k \cdot (1-x)^{n-k},
\end{equation}
where $C(n,k)$ is the binomial coefficient, $n$ is the number of aluminum ions in the nearest environment $(n = 0, 1, …, 6)$, and $x$ is the aluminum concentration.

Following to this approach, we estimate the change in the canting angles for the composition with $x = 0.05$. First, we estimate the canting angle due to DMI. Taking the magnetization value from our experiment as 0.1 \M\ (52~K) for pure \HFO\ and assuming that all of it results from the canting of the iron moment with a total magnetic moment $m_{0}^{Fe}$ = 5 \M, we obtain approximately $\beta_{\mathrm{DM}} = 1.15^\circ $. Returning to $x = 0.05$, we divide the magnetization into two contributions: $M_1 = P(0) \cdot \mathrm{sin}(\beta_{\mathrm{DM}}) \cdot m_{0}^{Fe}$, corresponding to configurations where the iron ions are surrounded by 6 iron neighbors, and $M_2 = P(1) \cdot \mathrm{sin}(\beta_{\mathrm{DM}} + \gamma_\mathrm{H}) \cdot m_{0}^{Fe} $, which takes into account the contribution from configurations with aluminum neighbors ions. 
The angle $\gamma_\mathrm{H}$ is the additional angle responsible for the increase in magnetization; $P(0)$ and $P(1)$ are coefficients determining the contributions of the combinations with 6 and 5 iron neighbors. By combinations we mean the statistical variants of the distribution of magnetic vacancies in the sample determined according to Eq.~\ref{eq:binom1}. For $x = 0.05$, we obtain $P(0) = 0.7351$. $P(1) = 6/7 \cdot 0.2649$, which includes combinations with one ($0.2321$) and more aluminum ions out of 6 possible neighbors. The factor 6/7 is the number of magnetic iron ions distributed among 7 possible positions. Thus, by introducing the $M_2$ addition to the total magnetization with an additional angle $\gamma_\mathrm{H}$ to the angle $\beta_{\mathrm{DM}}$, we describe the observed sharp increase in magnetization upon substitution in the iron subsystem. Calculating this contribution yields $\gamma_\mathrm{H} =  1.7668^\circ$.

At lower temperatures, the usual saturation of the magnetic moment and stabilization of the $\Gamma_2$ phase, as in the parent \HFO, are not observed; on the contrary, a rapid decrease in magnetization occurs. In work~\cite{vorob1989unusual}, the influence of the anisotropy constant $K_{ab}$ on the formation of the intermediate $\Gamma_{12}$ phase is discussed. As shown in~\cite{vorob1991orientational}, magnetic vacancies contribute to both anisotropy constants, saturating $K_{ab}$ faster, which leads to a larger deviation of the \textbf{G} vector and stabilization of the $\Gamma_1$ phase. 
However, based solely on our magnetization data, the $\Gamma_1$ phase cannot be unambiguously identified from $\Gamma_{12}$ and for that reason we label this phase as $\Gamma_{12}$ keeping in mind that it also might be a pure $\Gamma_{1}$.
Inside the $\Gamma_{12}$ phase observed in Fig.~\ref{M(T)}b, two regions can be distinguished, separated by a minimum at 40~K. At this point the deviation of the \textbf{G} vector stops and a reverse rotation process begins, leading to the stabilization of the $\Gamma_2$ phase. One can assume that this effect is caused by the enhanced competition of the rare‑earth ion anisotropies, because in the parent \HFO\ at temperatures below 30~K, the $K_{ab}$ anisotropy reaches saturation, while the increase of the $K_{aс}$ does not stop until the rare‑earth subsystem orders~\cite{vorob1989unusual}.

We were able to use the above approach to explain the mechanism of the magnetization change shown in Fig.~\ref{M(T)}(b) for several reasons. One of them is that the value of the weak ferromagnetic moment in the $\Gamma_{24}$ phases is the same as at low $T$ in $\Gamma_2$ phase, meaning that all moments of \Fe\ were polarized. The second is the monotonic dependence of $M(T)$ curve within the $\Gamma_4$ phase in this sample, while for $x = 0.1, 0.2$ it shows a broad maximum above the SRT temperature.
Thus, drawing the comparison with the parent \HFO\ we were able to associate the missing magnetization of $x = 0.05$ sample with statistical combinations described by $P(1)$ contribution.
However, for larger $x$ the application of such an approach seems unjustified, since the behavior of both factors in Figs.~\ref{M(T)}(c-d) is significantly different.

\section{CONCLUSION}

In this work, a series of \HFAO \ single crystals with $x = 0, 0.05, 0.1, 0.2$ were grown by the optical floating‑zone method, and their magnetic properties were investigated. A strong influence of magnetic vacancies, obtained by substituting iron with aluminum, on the temperature dependences of magnetization in the SRT region and on the N\'eel temperature was found.

Using the magnetization data, we constructed a phase diagram that demonstrates the change in the SRT structure as a function of aluminum concentration $x$. It shows that the introduction of magnetic vacancies leads to the appearance of a mixed $\Gamma_{12}$ phase in the substituted \HFAO\ below the SRT temperature, suppressing the $\Gamma_{2}$ phase observed in the parent \HFO. We believe that this behavior is explained by the competition between the $K_{ab}$ and $K_{ac}$ anisotropies, which in \HFO\ participate in the stabilization of the $\Gamma_1$ and $\Gamma_2$ phases, respectively.

\section{ACKNOWLEDGMENTS}

The study was supported by the Russian Science Foundation grant No. 26‑22‑20006, https://rscf.ru/project/26-22-20006/, and by the targeted funding (grant) of the Krasnoyarsk Regional Foundation of Science.

\bibliographystyle{apsrev4-2}  
\bibliography{HoFe1_xAlxO3}     

\end{document}